\documentclass[aps,prl,floatfix,twocolumn,showpacs,preprintnumbers,amsmath,amssymb,superscriptaddress]{revtex4}

\usepackage{graphicx}  
\usepackage{dcolumn}   
\usepackage{bm}        
\usepackage{stmaryrd}  
\usepackage{color}

\begin{document}

\preprint{APS/123-QED}

\title{Correlation of the angular dependence of spin-transfer torque and giant magnetoresistance in the limit of diffusive transport in spin valves }
\author{M. Gmitra}
\affiliation{Institute of Phys.,
             P. J. \v{S}af\'arik University, Park Angelinum 9,
             040 01 Ko\v{s}ice, Slovak Republic}
\author{J. Barna\'s}
\affiliation{Department of Physics, Adam Mickiewicz University,
             Umultowska 85, 61-614~Pozna\'n, Poland}
\date{\today}

\begin{abstract}
Angular variation of giant magnetoresistance and spin-transfer
torque in metallic spin-valve heterostructures is analyzed
theoretically in the limit of diffusive transport. It is shown
that the spin-transfer torque in asymmetric spin valves can vanish
in non-collinear magnetic configurations, and such a non-standard
behavior of the torque is generally associated with a
non-monotonic angular dependence of the giant magnetoresistance,
with a global minimum at a non-collinear magnetic configuration.

\end{abstract}

\maketitle

{\it Introduction:} The macroscopic model of electronic transport
in magnetic metallic multilayers, developed by Valet and Fert
\cite{Valet_93}, is commonly used for interpretation of
experimental results on current-perpendicular-to-plane (CPP) giant
magnetoresistance (GMR). The model includes several
phenomenological parameters which can be extracted from fitting
experimental and theoretical results on
resistance/magnetoresistance in collinear (parallel and
antiparallel) magnetic configurations. The model is based on the
assumption that the spin diffusion length is longer than mean free
path, and the latter is smaller than the layer thickness.
Recently, Penn and Stiles \cite{Penn_05} showed that the
Valet-Fert model is justified even for spin diffusion lengths
comparable to the mean-free paths. Moreover, the model fits well
to  experimental results even when the mean free paths are
comparable to the layer thicknesses.

Validity of the model, however, can be justified only when it
gives results which are in agreement with experiment not only in
the collinear configurations (too many fitting parameters), but
also in the full range of angles between magnetic moments of the
layers' magnetizations. This problem was not addressed yet. In a
recent paper \cite{Barnas_05} the Valet-Fert model has been
extended to noncollinear configurations, but only the angular
variation of the spin-transfer torque (STT) was analyzed there.
The spin torque plays a crucial role in the phenomenon of
current-induced magnetic switching (CIMS). It turned out that both
CPP-GMR and CIMS effects are correlated. Moreover, there are
normal and inverse GMR and also normal and inverse CIMS phenomena.
Accordingly, four different possibilities can be found in real
systems \cite{Barnas_05}. Indeed, the normal and inverse CPP-GMR
and/or CIMS have been demonstrated by manipulating the bulk and/or
interface spin-asymmetry parameters \cite{AlHajDarwish_04}.

It has been shown in Ref.\cite{Barnas_05} that the STT in
asymmetric structures can vanish at a noncollinear configuration,
which has a significant impact on the stability of magnetic
configuration. As a result, precessional states in zero magnetic
field were predicted in the Co/Cu/Py nanopillars
\cite{Gmitra_0607} and later experimentally confirmed
\cite{Boulle_07}. One may naturally expect that the nonstandard
behavior of STT may be associated with some anomalous angular
behavior of the CPP-GMR. This problem is addressed here, and we
show that a global minimum in resistance of asymmetric structures
may occur in a noncollinear configuration. This non-monotonic
behavior of the resistance (and consequently also GMR) is
generally accompanied by the non-standard angular dependence of
STT. These two features seem to be characteristic of asymmetric
systems in the diffuse transport regime, provided the system's
parameters obey some conditions.

{\it Model of CPP-GMR:} Within the diffusive approach
\cite{Valet_93,Barnas_05,Brataas_01}, spatial dependence of the
average electrochemical potential in a ferromagnetic (F) layer has
the general form; $\bar{\mu} = -\beta \, g + C x + G$, where the
axis $x$ is normal to the structure, and $g$ is the spin
accumulation, $g=A \exp(x/l_{\rm sf}) + B \exp(-x/l_{\rm sf})$,
with $l_{\rm sf}$ being the spin diffusion length. Similar
formula also holds for normal metal (N) layers, but with
$\beta=0$. All the constants ($A$, $B$, $C$, and others) entering
the general solution of the diffusive equations in different
layers can be determined from the corresponding boundary
conditions \cite{Valet_93,Barnas_05,Brataas_01}.

The driving field can be then calculated as $E(x) = (1/e)(\partial
\bar{\mu}/\partial x)$ \cite{Valet_93}. The presence of N/F
interfaces gives rise to additional voltage drops due to spin
accumulation in their vicinity. The total voltage drop can be then
written as $\Delta V = \sum_i \Delta V_i$, where $\Delta V_i$ is
the voltage drop in the $i$-th layer of the spin valve (voltage
drops at interface resistances will be included to the
ferromagnetic layers). When the index $i$ corresponds to a
ferromagnetic layer, $\Delta V_i = \Delta V_i^{\rm SI} + \Delta
V_i^{\rm spl}$. If, however, $i$ corresponds to a nonmagnetic
layer, $\Delta V_i = \Delta V_i^{\rm SI} + \rho_i d_iI_0$, where
$\rho_i$ is the bulk resistivity of the normal metal,  $d_i$ is
the corresponding layer thickness, and $I_0$ is the current
density. The voltage drops due to spin accumulation (in magnetic
and nonmagnetic layers) read
\begin{equation}\label{Eq:drop}
\Delta V_i^{\rm SI} = \int_{x\in d_i} [ E(x) - E_0 ]\,{\rm d}x,
\end{equation}
where the corresponding electric field $E_0$ is taken far from the
interface. Apart from this, $\Delta V_i^{\rm spl}=I_0\,R_i^{\rm
spl}$ for ferromagnetic films, $R_i^{\rm spl} = [(1/R_{i\uparrow})
+ (1/R_{i\downarrow})]^{-1}$, $R_{i\sigma} =  R_{i\sigma}^{\rm L}
+ d\rho_{i\sigma} +  R_{i\sigma}^{\rm R}$ (for $\sigma =\uparrow
,\downarrow$), with $\rho_{i\sigma}$ being the corresponding spin
dependent bulk resistivity, and $R_{i\sigma}^{\rm L}$
($R_{i\sigma}^{\rm R}$) denoting the interfacial resistances (per
unit square) associated with the left (right) interface of the
$i$-th (ferromagnetic) film.

The total resistance of the system (per unit square) is $R =
\Delta V / I_0$, while the magnetoresistance, $\Delta R(\theta) =
R(\theta)-R(0)$, describes a change in the total system resistance
when magnetic configuration varies from a noncollinear to parallel
one. We note that what one needs to calculate are the $\Delta
V_i^{\rm SI}$ contributions only. It is convenient to define
reduced magnetoresistance as
\begin{equation}
r(\theta) = \frac{R(\theta) - R_{\rm P}}{R_{\rm AP} - R_{\rm P}}\,.
\end{equation}
Several theoretical approaches have been proposed to describe the
angular variation of GMR \cite{Vedyayev_97}. The explicit form for
the reduced magnetoresistance, $r(\theta) = \sin^2(\theta/2) /
[1+\chi\cos^2(\theta/2)]$ has been extracted within the
magneto-circuit theory \cite{Kovalev_02} and diffusive approach
\cite{Shpiro_03}. In recent measurements on Py/Cu/Py valves
\cite{Urazhdin_05} the parameter $\chi$ has been treated as a
fitting parameter, and the $r(\theta)$ dependence was found to
describe the experimental data relatively well. However, this
formula breaks down for asymmetric spin valves, where a global
minimum of the system resistance may appear at a non-collinear
configuration \cite{Urazhdin_05}.

\begin{figure}[!h]
\includegraphics[width=0.8\columnwidth,angle=0]{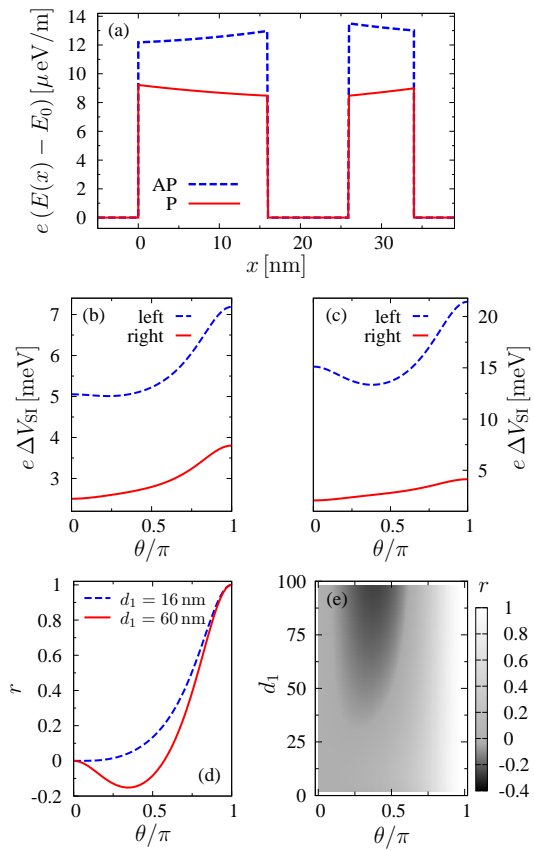}
\caption{(Color online) Transport characteristics of the biased
Cu/Co($d_1$)/Cu(10)/Co(8)/Cu spin valve. (a)~Spatial dependence of
the electric field in the system for $d_1=16\,{\rm nm}$ and for P
and AP configurations. Angular dependence of the voltage drops in
Co($d_1$) shown by dashed (blue) line and Co(8) by solid (red)
line for (b)~$d_1=16\,{\rm nm}$ and (c)~$d_1=60\,{\rm nm}$.
(d)~Angular dependence of the reduced magnetoresistance.
(e)~Reduced magnetoresistance as a function of $\theta$ and $d_1$.
In parts (a), (b) and (c) the current density $I_0=10^8\,{\rm
A/cm^2}$ was assumed. The other parameters are as in
Ref.\cite{Barnas_05}. } \label{Fig:MR}
\end{figure}

{\it Results:} The minimum in resistance (and also in GMR) in a
noncollinear configuration appears only in asymmetric spin valves.
In the following we discuss the angular dependence of the STT and
GMR in spin valves for positive current density, $I_0>0$
(electrons flow from right to left, or in other words charge
current flows from the layer of thickness $d_1$ to the layer of
thickness $d_2$). Figure \ref{Fig:MR}(a) shows electric field
profile in the Co($d_1$)Cu(10)Co(8) spin valve for $d_1=16\,{\rm nm}$
sandwiched between semi-infinite Cu leads and for both 
parallel (P) and antiparallel (AP) magnetic configurations. 
The angular dependence of the corresponding voltage drops within the 
Co layers is shown in Fig.\ref{Fig:MR}(b). The voltage drop within the 
Co(16) layer exhibits a very weak minimum for $\theta\simeq\pi/3$. The 
minimum becomes much more pronounced for larger layer thickness, as shown
in Fig.\ref{Fig:MR}(c) for $d_1=60\,{\rm nm}$. Since the total
voltage drop is a sum of all drops in the individual layers, the
GMR can exhibit a minimum at a non-collinear configuration when
the resistance decrease in the thick F layer overcomes the
resistance increase in the thin F layer. The global minimum arises
as a result of the spatial depletion of electrical field in the
thick F layer, which is a consequence of spin accumulation
discontinuity at the N/F interface controlled by the mixing
conductances. The reduced GMR, $r(\theta)$, is shown in
Fig.\ref{Fig:MR}(d) for both values of $d_1$. The non-monotonic
behavior of the reduced GMR is more pronounced in spin valves that
are more asymmetric, see Fig.\ref{Fig:MR}(e).

\begin{figure}[!h]
\includegraphics[width=0.95\columnwidth]{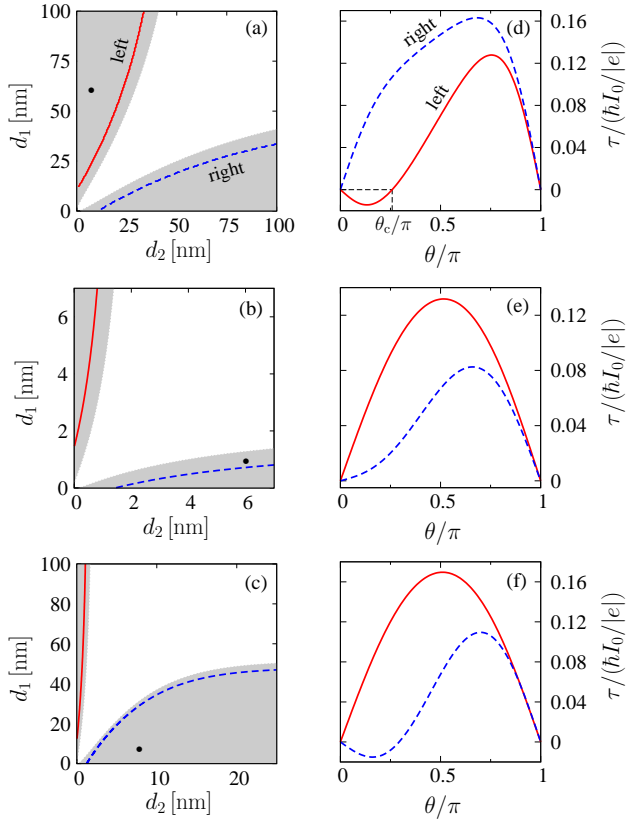}
\caption{(Color online) Diagrams illustrating presence of a global
magnetoresistance minimum at non-collinear configurations -- gray
regions -- and angular spin-transfer torque dependences. (a)~Diagram
for the Co($d_1$)Cu(10)Co($d_2$), (b)~Py($d_1$)Cu(10)Py($d_2$),
and (c)~Co($d_1$)Cu(10)Py($d_2$) spin valve. The solid (red) and
dashed (blue) lines denote critical thicknesses where
$\partial\tau/\partial\theta |_{\theta\to 0}=0$ for the torque
exerted on the left F layer of thickness $d_1$ and right layer
of thickness $d_2$, respectively. (d,e,f)~Angular dependence of
the spin-transfer torques acting on the left F layer of thickness
$d_1$ -- shown by solid (red) lines -- and right F layer of thickness
$d_2$ -- shown by dashed (blue) lines -- for systems corresponding
to the dots in the left panel~(a,b,c). }
\label{Fig:d-maps}
\end{figure}

In Figs \ref{Fig:d-maps}(a)-(c), the diagrams present the regions
of layer thicknesses, where the non-monotonic behavior of the
reduced GMR can be observed (gray regions). For the Co/Cu/Co spin
valves [Fig.\ref{Fig:d-maps}(a)] as well for the Py/Cu/Py ones
[Fig.\ref{Fig:d-maps}(b)], the diagrams are symmetric with respect
to $d_1=d_2$, and the non-monotonic angular variation of
the GMR (global minimum at a non-collinear configuration) can be
noticed for spin valves with significantly different layer
thicknesses. In Co/Cu/Py spin valves, where an additional asymmetry
appears due to different magnetic materials, a non-monotonic angular
variation of the GMR can be observed even for comparable layer
thicknesses, see Fig.\ref{Fig:d-maps}(c).
This is mainly due to strong asymmetry in spin diffusion lengths
of Co and Py, but difference in the bulk as well as interface spin
asymmetries of the Co and Py also contributes to the
non-monotonic behavior.

Experimental observations on Py/Cu/Py spin valves revealed a weak
non-monotonic angular variation of the GMR effect \cite{Urazhdin_05}. 
This has been attributed to the absorption of transverse spin accumulation 
in a non-collinear configuration, which reduces the resistance. 
Such absorption also gives rise to the STT acting on the F 
layer \cite{Stiles_02}, which in asymmetric
spin valves can exhibit an anomalous (non-standard) angular
dependence \cite{Gmitra_0607,Boulle_07}. In systems with a
non-standard STT, the transverse component of spin current
(accumulation) at the active N/F interface vanishes at a certain
noncollinear configuration. The presence of a GMR minimum in a
noncollinear configuration can be thus related to the non-standard
STT.

The STT appears due to absorption of the transverse spin current
component $\bm{j}_{\perp}$ at the N/F interface \cite{Stiles_02},
and can be calculated as
\begin{equation}
\bm{\tau} = \frac{\hbar}{2} \big( \bm{j}_{\perp}^{\rm L}
- \bm{j}_{\perp}^{\rm R} \big) \,,
\end{equation}
where the superscripts L and R denote the left and right
interface, respectively, associated with the F layer. Dependence
of the STT can also be expressed explicitly in terms of the mixing
conductances and spin accumulation at the normal-metal side of the
N/F interface \cite{Barnas_05,Gmitra_0607}. The STT consists
generally of both in-plane and out-of-plane components. Since the
latter component is much smaller than the former one (due to small
imaginary part of the mixing conductances \cite{Xia_02}), in the
following discussion we will consider only the in-plane component.
In asymmetric spin valves, the proper choice of magnetic materials
and/or layer thicknesses can result  in  vanishing  STT at a
non-collinear magnetic configuration \cite{Gmitra_0607}. Such a
non-standard STT destabilizes both collinear configurations for
positive current and stabilizes both configurations for negative
current \cite{Barnas_05,Gmitra_0607}. The former case is of
particular interest as the non-standard torque leads to
current-induced steady state oscillations in the absence of
external magnetic field \cite{Gmitra_0607,Boulle_07}. In
Fig.\ref{Fig:d-maps}(d) we show the angular dependence of STT in
the Co(60)Cu(10)Co(8) spin valve exerted on the Co(60) (solid
line) and Co(8) (dashed line). The STT acting on the Co(8) layer
destabilizes P and stabilizes AP configuration, whereas the torque
acting on the Co(60) vanishes at a non-collinear configuration and
stabilizes both P and AP configurations. The torques in
Fig.\ref{Fig:d-maps}(d) correspond to the system indicated by the
dot in Fig.\ref{Fig:d-maps}(a). This point is below the critical
line, given by $\partial\tau/\partial\theta|_{\theta\to 0}=0$,
which identifies the region where a non-standard STT acting on the
Co($d_1$) layer appears. When the layer thicknesses are  above the
critical line, but still in the gray region, see the dot in
Fig.\ref{Fig:d-maps}(b), the torque acting on the particular F
layer vanishes only for the collinear configurations, as shown in
Fig.\ref{Fig:d-maps}(e) for the Py(1)Cu(10)Py(6) spin valve, but
reduced GMR still exhibits a global minimum for a non-collinear
configuration. Since the critical lines are close to the boundary
of the non-monotonic angular GMR behavior (gray regions), the
non-standard STT is correlated with the non-monotonic angular
variation of GMR.

\begin{figure}[!h]
\includegraphics[width=0.95\columnwidth]{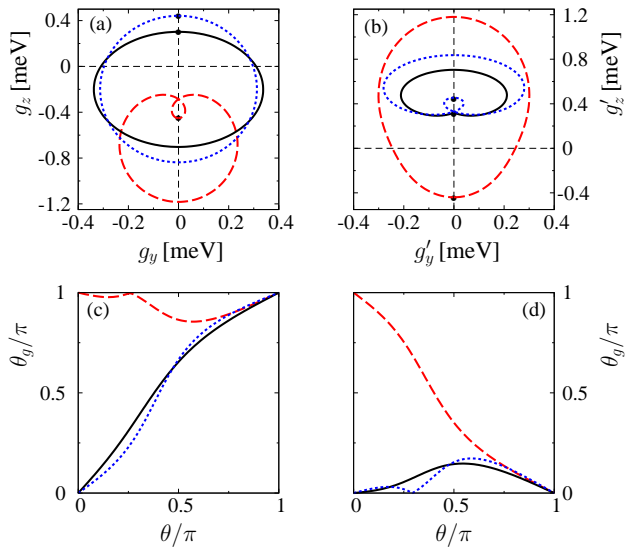}
\caption{ (Color online) Parametric plots of the in-plane spin
accumulation components and angular dependences of the angle
$\theta_g$ between the spin accumulation and spin moments.
(a)~spin accumulation at the left (F/N) interface; (b)~spin
accumulation at the right (N/F) calculated at the normal metal
side (N) in the vicinity of the active interfaces. The spin
accumulation components are expressed in their local reference
frames. The dashed (red) lines correspond to Co(60)Cu(10)Co(8);
solid (black) lines to Py(1)Cu(10)Py(6); and dotted (blue) lines
to Co(8)Cu(10)Py(8). The filled dots correspond to the parallel
configuration of the layer magnetizations. Angular dependence of
the $\theta_g$ at the (c)~left F/N and (d)~right N/F interface. }
\label{Fig:polar}
\end{figure}

At the critical angle $\theta_{\rm c}$, where the torque $\tau$
vanishes, the transverse component of spin accumulation at the
active interface disappears. In a general case, however, the angle
$\theta_g$ between the spin moment of the F layer and spin
accumulation vector at the normal-metal side of the N/F interface
is nonzero. Angular dependence of the STT can be then expressed as
a function of $\theta_g$ \cite{Barnas_05}. In
Fig.\ref{Fig:polar}(a) and (b) the in-plane spin accumulation
components at the normal-metal side (in the spacer layer) at the
left and right interfaces are shown for the spin valves considered
in Figs~\ref{Fig:d-maps}(d)-(f). The components are expressed in
local coordinate frames, where $g_z$ points along magnetization in
the left F layer whereas $g^{\prime}_z$ along magnetization in the
right F layer. The STT acting on Co(8) in the Co(8)Cu(10)Py(8)
spin valve exhibits regular behavior. Spin accumulation in the P
configuration is positive and has comparable amplitudes in the
vicinity of both interfaces [see the dots on the dotted lines in
Figs~\ref{Fig:polar}(a) and (b)] due to long spin-flip length in
Cu ($l_{\rm sf}\simeq 1\,{\rm\mu m}$). When magnetization of the
right layer rotates in the film plane, spin accumulation at the
left interface roughly follows the net-spin of the Py(8) layer,
and the angle $\theta_g$ is a monotonic function of $\theta$, see
Fig.\ref{Fig:polar}(c). Figure \ref{Fig:d-maps}(f) shows that STT
exerted on the Py(8) layer vanishes in a non-collinear
configuration ($\theta=\theta_{\rm c}$), for which the
$g^{\prime}_y$ component also vanishes. At $\theta=\theta_{\rm c}$
one finds $\theta_g=0$. For the Co(60)Cu(10)Co(8) system, STT
acting on the Co(60) layer shows a non-standard behavior which is
qualitatively similar to that for the Py(8) layer in the
Co(8)Cu(10)Py(8) spin valve. Angular variation of STT for the
Py(1)Cu(10)Py(6) spin valve, shown in Fig.\ref{Fig:d-maps}(e),
vanishes regularly in P and AP configurations, and $\theta_g$ is a
monotonic function of $\theta$.

Non-collinear configuration of the F layer magnetizations leads to
discontinuities of the spin accumulation at the F/N interfaces
[angle $\theta_g$ in Figs 3(c) and (d)]. From this we deduce that 
if one takes the thickness of one of the F layers smaller than the 
corresponding spin diffusion length and thickness for the second F layer 
is larger than the appropriate spin diffusion length, then the spin 
accumulation is predominately determined by the later F layer. One finds
then non-standard STT and non-monotonic GMR angular behavior. We
have found that this behavior is mostly controlled by the mixing
conductance of the interface between spacer layer and that F layer
whose thickness is smaller than the corresponding spin diffusion
length. For instance reducing the mixing conductance  at the
Co(8)/Cu(10) interface in the Co(8)Cu(10)Py(8) valve by about 50\%
lifts the non-standard STT and GMR behavior.

In conclusion, what stems from the above results is a need for
further experimental investigations, and that Co/Cu/Py system is a
good candidate to test the theoretical predictions. This also
could answer the question whether the diffusive approach used to
analyze CPP-GMR in collinear configurations is well justified. To
arrive at more convincing conclusions one also should correlate
the results on GMR with those on STT.

{\it Acknowledgments:} We thank A.~Fert and V. Cros for useful
discussions. This work was partly supported as a research project
MVTS POL/SR/UPJS07, VEGA 1/0128/08, and by EU through MC-RTN
SPINSWITCH (MRTN-CT-2006-035327). J.~B. also acknowledges support
by funds from the Polish Ministry of Science and Higher Education
as a research project in years 2006-2009, and by the Polish National
Scientific Network ARTMAG: Magnetic nanostructures for spintronics.

\end{document}